\documentclass[12pt]{iopart}

\usepackage{graphicx}
\usepackage{iopams}  
\usepackage{accents}
\begin{document}

\title{Evolution of non-interacting entropic dark energy and it's phantom nature.}

\author{ Titus K. Mathew, Chinthak Murali and Shejeelammal J}

\address{Department of Physics, Cochin University of Science and Technology, Kochi, India}
\ead{titus@cusat.ac.in}
\begin{abstract}
Assuming the form of the entropic dark energy as arises form the surface term in the Einstein-Hilbert's action, it's evolution were 
analyzed in an expanding flat universe. The model parameters were evaluated by constraining model using the Union data on Type Ia supernovae. We found that 
in the non-interacting case, the model predicts an early decelerated phase and a later accelerated phase at the background level. The evolution of the Hubble parameter, dark energy density,
equation of state parameter and deceleration parameter were obtained.  The model is hardly seems to be supporting the 
linear perturbation growth for the structure formation. We also found that the entropic dark energy 
shows phantom nature  for redshifts $z<0.257.$ During the phantom epoch, the model predicts big-rip effect at which both the scale factor of 
expansion and the dark energy density 
become infinitely large and the big rip time is found to be around 36 Giga Years from now.
\end{abstract}

\maketitle

\section{Introduction}
\label{intro} The mounting observational evidences\cite{Perl1,Riess1} have proved that the current universe is accelerating in expansion and the acceleration was began in the recent past in the history 
of the universe. It was conjectured that this acceleration is caused by the exotic form of energy called dark energy (DE) which can able to produce negative pressure, unlike
the conventional form of matter. The origin and evolution of dark energy is still the most important problem for the physicist today. The standard $\Lambda$CDM model 
in which the universe consists of cod dark matter and cosmological constant as dark energy is matching well
with the observational data, but it is facing the severe cosmological constant
problem that the theoretical value of the dark energy density 
as predicted form the field theoretical considerations is nearly 120 orders larger than the observational value \cite{Sahni1,Paddy1}. This motivates the 
consideration of the dynamical DE models. Various DE models were proposed aiming mainly on solving the 
cosmological constant problem. For a review of dark energy models, see, Copeland et. al. \cite{Copeland1}, Sahni et. al. \cite{Sahni1} and references therein.
Among the different classes of models the quintessence models \cite{Cald1,Peebles1} and K-essence \cite{Chiba1} models based on the scalar fields, while Chaplygin gas 
model\cite{Alam2} is an example for the one trying unify both dark matter and DE. Various other models are existing in the current literature, for instance, 
holographic dark energy models\cite{Li1}, Agegraphic\cite{Cai1} models etc. Among these variety of models, entropic-force DE model gained much attention 
recently. Entropic DE (EDE)  model 
was first proposed by Easson et. al.\cite{Easson1,Easson2}. This model is based on the idea of entropic gravity, proposed by Verlinde\cite{Verlinde1}. Verlinde 
explained gravity as a kind of force related to the change in entropy, hence the name entropic gravity, and he derived the field equations of gravity 
from the second law of thermodynamics. The entropic dark energy in the entropic gravity model is being arised from the surface term, often neglected,
 appearing in the Einstein-Hilbert's action. It was conjectured that the horizon, the boundary of the universe is acting as the so called `holographic screen'.

In the work of Easson et.al., it was shown that surface part in the action would add a positive term to the acceleration equivalent to 
 $C_H H^2 + C_{\dot{H}} \dot{H},$ where $H$ is the Hubble parameter, $C_H \, \& \, C_{\dot{H}}$ are the model parameters assumed to be in the range 
$3/2\pi \leq C_H \leq 1$ and  $ 0\leq C_{\dot{H}} \leq 3/2\pi,$ where over-dot represents a derivative with time.
They have demonstrated from error plot analysis using Type Ia supernovae data, that, the model is in good agreement in predicting 
the distance modulus of various supernovae and the universe moves smoothly from a decelerating to an accelerating epoch at around a redshift $z\sim 0.5.$
Later this model was considered by 
Spyros et. al.\cite{Spyros1}, as a running vacuum, with constant equation of state, -1 so that  the model leads to 
either eternal acceleration or eternal deceleration for the respective parametric range hence doesn't predict a transition from an early  decelerated phase to a latter 
accelerated phase of expansion and hence they argued that EDE model is not viable both 
at the background level and perturbation level.  For obtainig this result they have assumed a general conservation law, which implies a energy transffer 
between the EDE and other cosmic components.  On the other hand 
 Easson et al. doesnt assume any energy transffer, hence each component assumes sepertate conservation law.

In the present work we have assumed the general form for the EDE density and derived the Hubble parameter by following the Easson et al. approach, 
where there is no interaction between the dark energy and other cosmic componens, especially matter. The model is then contrasted with the 
observational data  on Type Ia supernovae to extract the model parameters. The various cosmological parameter like dark energy density, equation of state,
deceleration parameter were calculated and their evolution are studied.
The paper is organized as follows. In section two we obtained the Hubble parameter and constraint the model with observational data to extract the model 
parameters. The error bar plots with supernovae data were constructed and with the model prediction. The model is also compared with the Stern et al. data on 
Hubble parameter at various redshifts. Section three consists of our analysis of the various cosmological parameters and also 
results on the $Om$ diagnostic analysis of the model. In section 
four we have presented our observations on the big rip singularity occurred in this model followed by conclusions in section five.

\section{Entropic dark energy, Hubble parameter and estimation of model parameters}
\label{sec:2} The surface term in the action, which is often neglected, will causes the EDE. The Einstein Hilbert action including the surface term can be 
schematically expressed as\cite{Easson1,Hawking1}
\begin{equation}
 I = \int_M (R + L_m) + \frac{1}{8\pi} \oint_{\partial M} K
\end{equation}
where $R$ is the scalar curvature, $L_m$ is the Lagrangian corresponds to matter, filed and $K$ is the trace of the extrinsic curvature of the boundary. The variation 
of this action  with respective boundary contribution will lead to the Einstein's field equation, which when combined with 
Friedmann metric, gives the Firedmann equations, of which especially the acceleration equation become\cite{Easson1},
\begin{equation}
 {\ddot{a} \over a} = -\frac{4\pi G}{3}\left(\rho + 3 P\right) +\left( C_{\dot{H}} \dot{H} + C_H H^2 \right)
\end{equation}
In this equation the last two positive terms on the right hand side constitute the entropic dark energy density, with coefficients $C_{\dot{H}}$ and 
$C_H,$  which constitute the EDE density,
\begin{equation} \label{eqn:rhode}
 \rho_{EDE}=3\left(C_H H^2 + C_{\dot{H}} \dot{H}\right)
\end{equation}
where $C_H$ and $C_{\dot{H}}$ are the model parameters and we follow the natural system of universe, $8\pi G=c=1.$ For a spatially flat Friedmann universe, which is 
being favored by cosmic microwave background (CMB) observations and predicted by inflationary models \cite {Bernardis1,Hanany1}, this DE satisfies the Friedmann equation,
\begin{equation}\label{eqn:H1}
 H^2 = \frac{\rho_m}{3} + \frac{\rho_{EDE}}{3}
\end{equation}
where $\rho_m$ is the matter or effectively cold dark matter density. and we use the natural system unit in which $8\pi G=c=1$. Assuming no interaction between 
dark energy and non-relativistic matter, the 
matter and entropic DE satisfies the conservation equation separately,
\begin{equation}\label{eqn:con1}
 \dot{\rho}_m + 3 H \left(\rho_m+P_m\right)=0
\end{equation}
\begin{equation}\label{eqn:con2}
 \dot{\rho}_{EDE}+3 H \left(\rho_{EDE}+P_{EDE}\right)=0
\end{equation}
where $P_{m/EDE}$ are the pressures of the matter/EDE components of the universe.

Equations (\ref{eqn:rhode}) and (\ref{eqn:H1}) were properly combined to obtain the differential equation for Hubble parameter evolution,
\begin{equation}
 {dh^2\over dx}+\left({2(C_H -1) \over C_{\dot{H}}}\right) h^2 + {2 \Omega_{m0} \over C_{\dot{H}}} e^{-3x}=0
\end{equation}
where $h=H/H_0, \, \Omega_{m0}$ is the present value of mass parameter of dark matter and $x=\log a$ with $a$ as the scale factor of expansion. The solution
of this will gives the evolution of Hubble parameter as,
\begin{equation}\label{eqn:hsol1}
 h^2(x)= \eta e^{-3x} + \left(1 - \eta \right) e^{-(2(C_H -1)/C_{\dot{H}}) x}
\end{equation}
where,
\begin{equation}\label{eqn:eta}
 \eta= \left({\Omega_{m0} \over 1+(\frac{3}{2}C_{\dot{H}} - C_H)}\right),
\end{equation}
which shows that $\eta=\Omega_{mo}$  if $C_H=3C_{\dot{H}}/2$, $\eta<\Omega_{m0}$ if $C_H<3C_{\dot{H}}/2$ and $\eta>\Omega_{m0}$ if $C_H>3C_{\dot{H}}/2.$ 
The Hubble parameter 
given by the above equation also satisfies the condition $h=1$ for $z=0$ corresponds to $x=0,$ apt for the current state of the universe. Especially for the case $C_H=3C_{\dot{H}}/2$ the 
Hubble parameter reduces to,
\begin{equation}
 h^2=\Omega_{m0} e^{-3x}+ \left(1-\Omega_{m0}\right) e^{-(3(C_H-1)/C_H)x},
\end{equation}
and with additional constraint $\Omega_{m0}=1,$ corresponds to the matter dominated case the Hubble parameter become,
\begin{equation}
 h^2 = e^{-3x},
\end{equation}
which in fact the Einstein-de Sitter model. 
\subsection{Contrasting with Type Ia data}
We have evaluated the model parameters $C_H, C_{\dot{H}}$ and also the present value of Hubble parameter, $H_0$ by constraining model 
using the observation data. These parameters are determined by minimizing the $\chi^2$ function,
\begin{equation}
 \chi^2 = \sum_i {\left[\mu_{obs}(z_i) - \mu_{th}(z_i) \right]^2 \over \sigma_i^2},
\end{equation}
where $\mu_{obs}(z_i)$ is the observed distance modulus of the supernova at redshift $z_i$, $\mu_{th}(z_i)$ is the corresponding theoretical modulus and 
$\sigma_i$ is the uncertainty in the SNe observation.
We have used the ``Union'' SNe data set\cite{Kowalski1}, which consists 307 type Ia supernova. The theoretical distance modulus is,
\begin{equation}
 \mu_{th}(z_i,\alpha,\beta,H_0)=5 \log_{10} \left({d_L \over Mpc} \right) + 25,
\end{equation}
where $d_L$ is the luminosity of the supernova and is given by,
\begin{equation}\label{eqn:dL}
 d_L(z_i,\alpha,\beta,H_0)=c (1+z) \int_0^{z_i} {dz^{'} \over H(z^{'},\alpha,\beta,H_0)},
\end{equation}
where $c$ is the light speed. The best estimates of the parameters $C_H, C_{\dot{H}}$ and $H_0$ for three different values of $\Omega_{m0}=0.25, 0.27, 0.30$ 
are obtained by minimizing the $\chi^2$ function and are shown in 
table \ref{table:t1}.
\begin{table} 
\centering
\begin{tabular}{|c|c|c|c|c|c|}
\hline
 Model & $\chi_{min}^2$ & $\chi_{min}^2$/d.o.f & $\alpha$ & $\beta$ & $H_0$ \\ \hline
EDE with $\Omega_{m0}=0.30$ & 311.29 & 1.024 & 0.813 & 0.415 & 70.59 \\ \hline  
EDE with $\Omega_{m0}=0.27$  & 311.29 & 1.024 & 0.832 & 0.374 & 70.59 \\ \hline 
EDE with $\Omega_{m0}=0.25$  & 311.29 & 1.024 & 0.844 & 0.346 & 70.59 \\ \hline
$\Lambda$CDM model & 311.93 & 1.026 & - & - & 70.03 \\ \hline
\end{tabular}
\caption{Best estimates of the parameters using supernovae data for different  $\Omega_{m0}.$  The term, d.o.f = degrees of freedom = N - n, N=307, the number of data points, 
n=3, number of parameters in the model.  For comparison we have also evaluated the corresponding values for 
the standard $\Lambda$CDM model for the same data set.}
\label{table:t1}
\end{table}
The $\chi^2_{min}$=311.29 and the Hubble parameter $H_0=70.59$  are the same for the three values of the dark mass parameter but $C_H$ parameter value decreases 
as $\Omega_{m0}$ increases and $C_{\dot{H}}$ parameter value increases with $\Omega_{m0}.$ For a 
comparison we also evaluated the $\chi^2_{min}$ for the $\Lambda$CDM model, which give values almost close to the EDE values.
\begin{figure}[h]
 \centering
\includegraphics[scale=0.7]{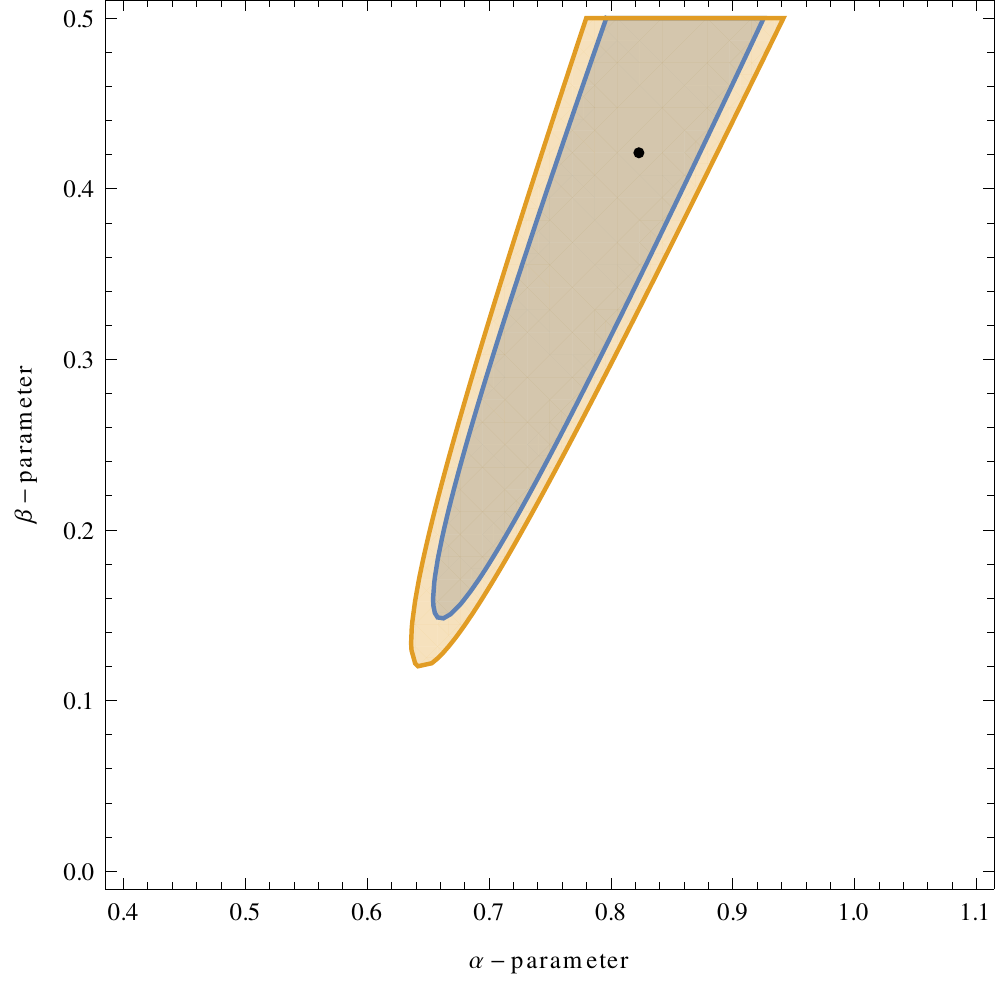}
\caption{Confidence intervals for the parameters $\alpha \equiv C_H$ and $\beta \equiv C_{\dot{H}}$ for the Hubble parameter $H_0=H(z=0)=70.42$ and $\Omega_{m0}=0.3$
 using the Union supernovae data. The confidence intervals 
shown are corresponds to 99.73$\%$ probability (inner one) and 99.99$\%$ probability (outer one). The dot represents the values of $(C_H,C_{\dot{H}})$ corresponds 
to $\chi^2_{min}.$ }
\label{fig:contour1}
\end{figure}
The statistical correction for the parameters values were obtained by constructing the confidence regions in figure \ref{fig:contour1} for $\Omega_{m0}=0.3.$ 
The best fit corresponds to 99.73$\%$ are $C_H=0.813 \pm 0.056, C_{\dot{H}}=0.415\pm 0.061$ and for 99.99$\%$ probability are $C_H=0.813\pm0.072, C_{\dot{H}}=0.415\pm 0.079.$ 

The best estimated values shows that $2(C_H-1)/C_{\dot{H}} \simeq -0.902,$ which implies that 
at large redshift, $z>>1,$ the Hubble parameter become,
\begin{equation}
 h^2 \simeq \eta (1+z)^3,
\end{equation}
ensures that the early universe has gone through a matter dominated phase. 
In comparison with the corresponding equation in standard $\Lambda$CDM model, i.e. $h^2=\Omega_{m0}(1+z)^3,$ the parameter $\eta$ is now corresponds to 
the dark mass parameter $\Omega_{m0}$ in $\Lambda$CDM model, however it is slightly large, $\eta=1.23\Omega_{mo},$ as per the best estimates of the model parameters. 
 This hike is 
due to the fact that the EDE is mimicking the dark matter characteristic in the early matter dominated epoch, a fact to be clear from the 
considerations of equation of state parameter, and is 
contributing towards the total non-relativistic mass density at $z>>1$.
Similar situations of increased mass parameter were discussed in many recent literature, for instance the mass parameter $\tilde{\Omega}_{m} =1.1 \Omega_{m0}$ in the quintessence model 
considered in reference\cite{Sahni6}. However for the sufficient growth of perturbations in the matter dominated epoch of a universe consists of quintessence filed, 
it is argued that 
$\tilde{\Omega}_{mo} \lesssim 1.15 \Omega_{m0}$ \cite{Starobinsky1}. 
Taking this constraint in to consideration, it seems that the non-interacting EDE model hardly supporting the perturbation 
growth, required for the structure formation in the matter dominated era, hence it may not 
be viable at the perturbation level.

\begin{figure}
 \centering 
\includegraphics[scale=0.7]{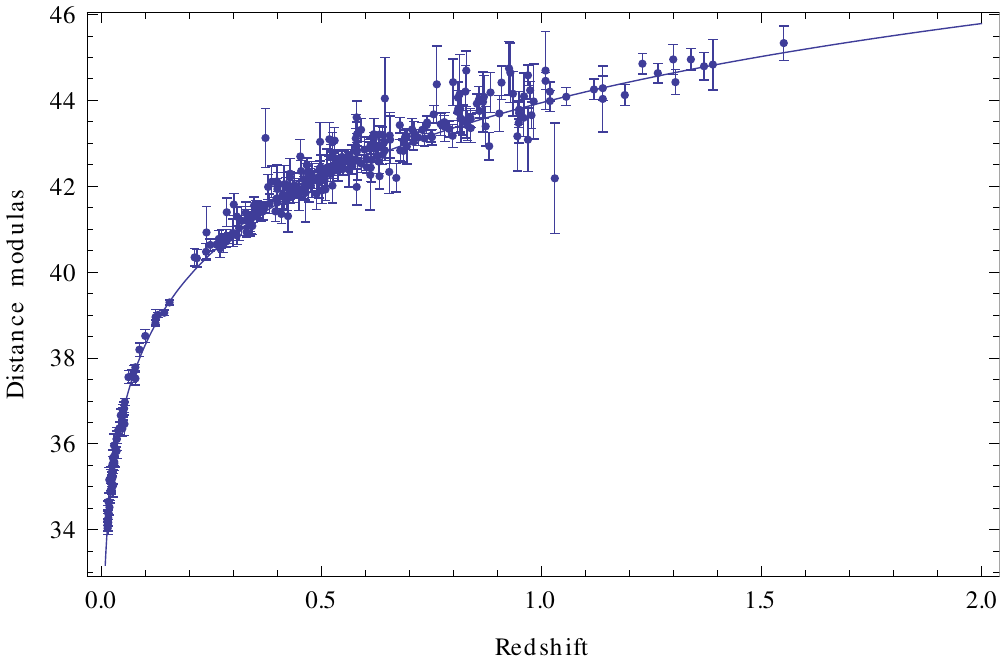}
\caption{Comparison of the distance modulus between the EDE model with $\Omega_{m0}=0.3$ and supernovae data. The error bars are corresponds to the observational data and the continuous line 
is the prediction from the present model.}
\label{fig:dL}
\end{figure}

\begin{figure}
 \centering
\includegraphics[scale=0.7]{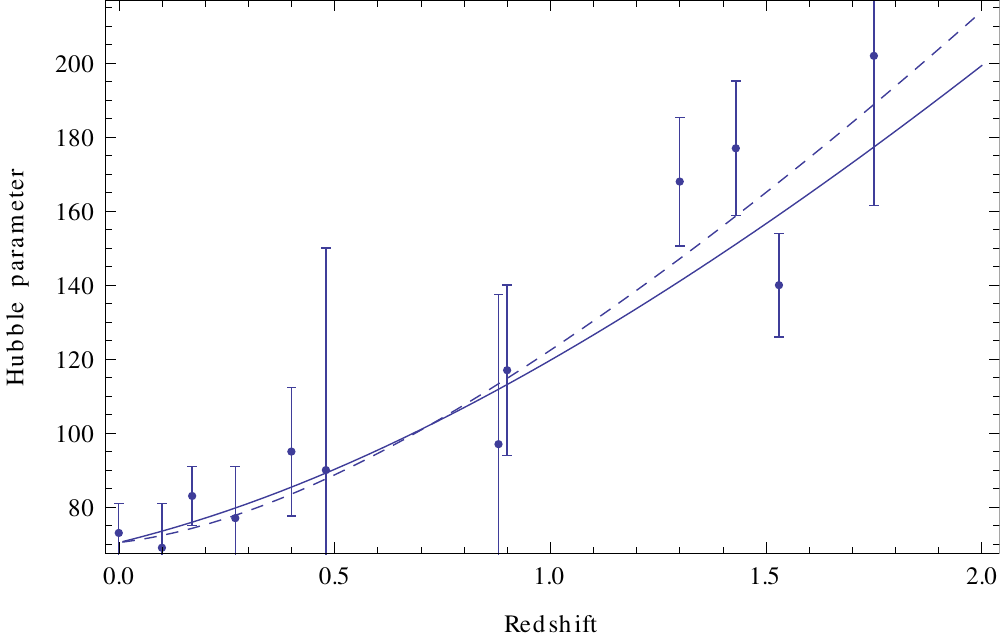}
\caption{Hubble parameter evolution of EDE model with $\Omega_{m0}$ compared with Stern data on Hubble parameter and $\Lambda$CDM model. the error bar lines are corresponds to Stern data, dotted 
line corresponds to Hublle parameter evolution of EDE model and continuous line corresponds to $\Lambda$CDM model.}
\label{fig:hubble-stern1}
\end{figure}

Using the best estimates of the model parameters, we have compared the EDE's prediction of the distance modulus $d_L,$ of supernovae at various redshift
 in equation (\ref{eqn:dL}) with 
supernovae data and is shown in 
figure \ref{fig:dL}.
The figure shows good agreement between theory and observation at low redshifts.
Similarly the Hubble parameter evolution in EDE model is compared with the Stern data\cite{Stern1} and also with the $\Lambda$CDM model in figure \ref{fig:hubble-stern1}.
It is seen that the Hubble parameter evolution of the EDE model is fitting fairly well with the observational data at lower redshift but high redshift the fitting is not 
seems to be good. 
At the low red shift region for the expansion rate in the EDE model is slightly lagging behind the the $\Lambda$CDM model, at around $z \sim 0.7$ the expansion rate 
in the non-interacting EDE model making cross-over into large rate and 
is very well leading  $\Lambda$CDM model at large redshift.
This is due to the fact that at very large redshift redshift the EDE dark energy is almost having the behavior of cold dark matter.
While discussing the model independent extraction of dark energy properties,
Sahni et. al. \cite{Sahni3} shows a similar type of contrast in the behavior of the Hubble parameter between the model independent dark energy and $\Lambda$CDM model.

\section{Evolution of cosmological parameters}
\label{sec:8}In the present model, the dark energy density evolution with respect to redshift can be obtained form equation (\ref{eqn:hsol1}) as,
\begin{equation}\label{eqn:deden1}
\Omega_{EDE}=\eta e^{-3x} + \left(1 - \eta \right) e^{-(2(C_H -1)/C_{\dot{H}}) x} - \Omega_{m0} e^{-3x}.
\end{equation}
In terms of the scale factor the above equation become,
\begin{equation}
 \Omega_{EDE}=(\eta  - \Omega_{m0}) a^{-3} + \left(1 - \eta \right) a^{-(2(C_H -1)/C_{\dot{H}}) }.
\end{equation}
For the best estimates of the model parameters, it can be easily seen that $2(C_H-1)/C_{\dot{H}} \simeq -1$, so the last term in the above equation 
become proportional to $a^1.$  Hence in the remote past of the universe (at large redshift, $z>>1$) corresponds to $a \to 0$ 
 the EDE density become, $\Omega_{EDE} \simeq (\eta - \Omega_{m0})a^{-3},$ 
which corresponds to the behavior similar to the cold dark matter. But at small redshift corresponds to large $a$ 
the dark energy density, $\Omega_{EDE} \sim \left(1 - \eta \right) a,$ grows almost in proportion to the scale factor. 

\begin{figure}
 \centering
\includegraphics[scale=0.7]{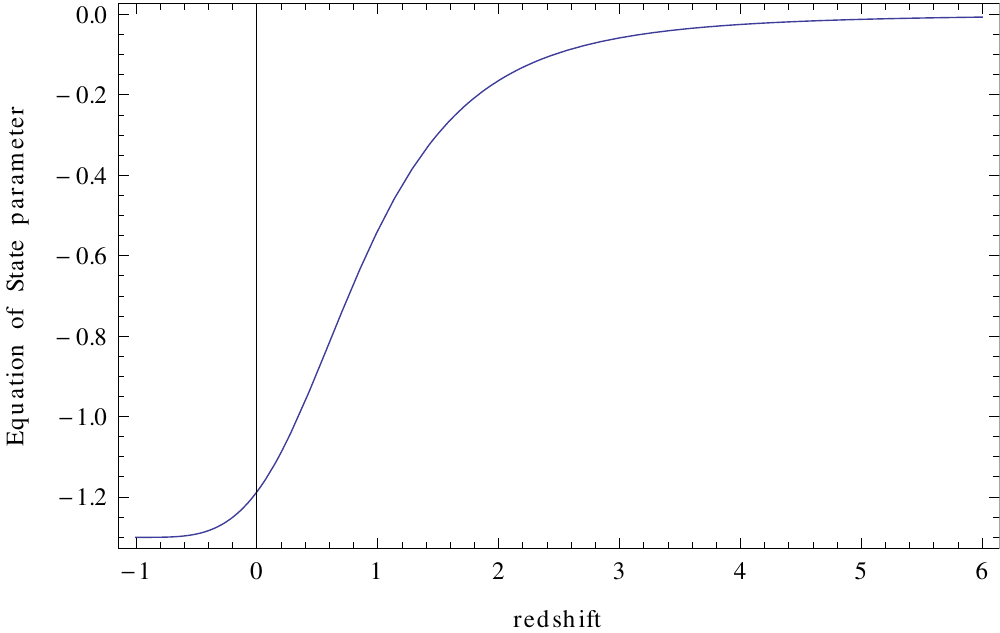}
\caption{Evolution of the equation of state parameter for the best fit values of the parameters.}
\label{fig:eos1}
\end{figure}

The equation of state parameter can be obtained from standard 
relation\cite{mathew1} in terms of redshift as,
\begin{equation} \label{eqn:eosp}
 \omega_{EDE}=-1 + \frac{1}{3} \left({3(\eta - \Omega_{m0})(1+z)^3 + ({2(C_H - 1)\over C_{\dot{H}}}) (1 - \eta) (1+z)^{2(C_H-1)/C_{\dot{H}}} \over (\eta - \Omega_{m0})
(1+z)^3 + (1 - \eta) (1+z)^{(2(C_H-1)/C_{\dot{H}}} } \right)
\end{equation}
At very large redshift, $z>>1$ only the terms proportional to $(1+z)^3$ will effectively contribute, and other terms will be negligibly small, as a result 
$\omega_{EDE} \to 0.$ That confirms that in very early stage the EDE has a behaviour similar to the cold dark matter.
But as $z \to -1$ the equation of state will behaves as, $\omega_{EDE} \to -1+2(C_H-1)/C_{\dot{H}},$ 
which is for the best estimates of 
parameters (for $\Omega_{m0}=0.3$),  become, $\omega_{EDE}(z\to-1) \to -1.3.$
 The evolution of the equation of state 
parameters given figure \ref{fig:eos1}. 
The figure shows that EDE equation of state parameter crosses the phantom divide, $\omega_{EDE}=-1,$ and proceed towards still lower values in the latter evolutions,
before stabilizing at around $-1.3.$ During it's phantom behavior at which the equation of state is less that -1, 
the EDE violates the null energy condition, $(\rho+p)>0$ and the energy density increase 
during the further evolution. 
In 
fact we have already seen form the previous paragraph that the EDE density does shows such a behavior in the late universe.
From literature one can find that phantom behavior can be constructed from scalar field with a negative kinetic term,
\begin{equation}
 \rho_{DE}=-\frac{1}{2}\dot{\phi}^2 + V(\phi) \\
p_{DE}=-\frac{1}{2}\dot{\phi}^2-V(\phi)
\end{equation}
which will leads to 
\begin{equation}
 \omega_{DE}={-\frac{1}{2}\dot{\phi}^2-V(\phi) \over -\frac{1}{2}\dot{\phi}^2+V(\phi)}.
\end{equation}
Where $\phi$ is the scalar field, $V(\phi)$ is the potential energy and the above equation gives $\omega_{DE}<-1$ if $\dot{\phi}^2>0.$
The current value of the equation of state in the present model is noted as $\omega_{EDE}(z=0) \sim -1.185.$ This values is higher than the observationally constrained value, 
around $\omega_{EDE}(z=0) \sim -0.93,$ \cite{Komatsu1} obtained 
form the joint analysis of WMAP+BAO+SNe data. The present value of the equation of state parameter of this model indicating the phantom nature of the entropic 
dark energy in the current phase of the universe. It is to be noted that some latest results on the equation of state in fact prefer phantom dark energy models.
For example
one of the latest cosmological data gives, $\omega_{DE}=1.04^{+ 0.09}_{-0.10}$\cite{Nakamura1}, while in reference \cite{Komatsu1}, the equation of state parameter 
is deduced as $\omega_{DE}=-1.10^{+0.14}_{-0.14}.$ The recent results from Plank satellite gives, $\omega_{DE}=-1.49^{+0.65}_{-0.57}$ \cite{Plank1}.

A forecast of the phantom behavior of dark energy is the occurrence of big-rip effect or future singularity\cite{Caldwell1,Nijori1,Nijori2}, due to the tremendous increase
in the dark energy density as the universe expands. We have already shown that the density is increasing in 
proportion to the scale factor at sufficiently lower redshifts. So the big rip is seems to be inevitable in this model. We will analyze this fact in a later section.

\begin{figure}
 \centering 
\includegraphics[scale=0.7]{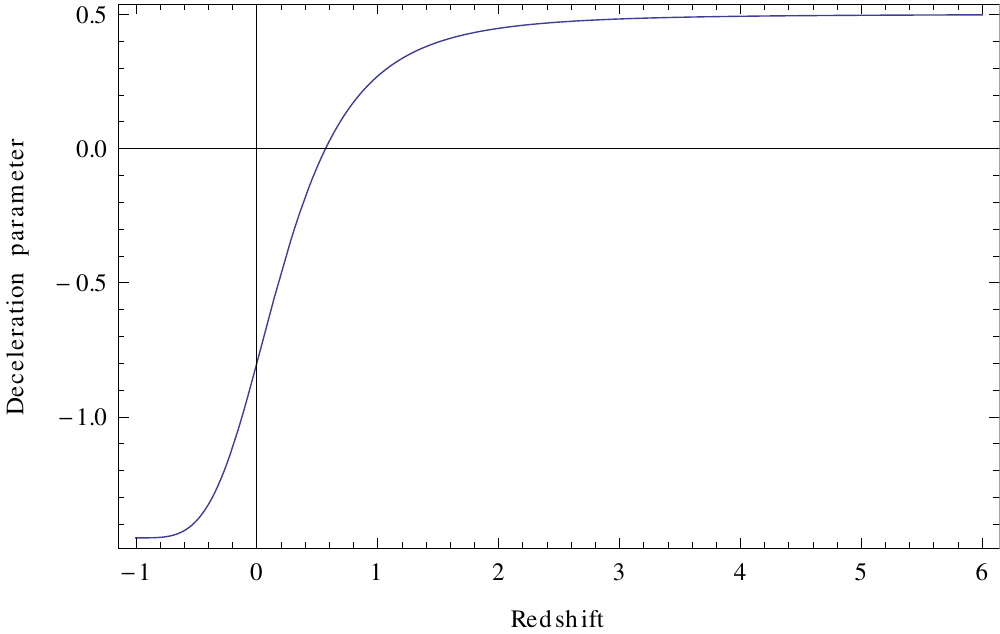}
\caption{The evolution of the deceleration parameter with redshift.}
\label{fig:decp}
\end{figure}

The deceleration parameter can be obtained using the relation,
\begin{equation}
 q = -1 -\frac{1}{2} {dh^2 \over dx}.
\end{equation}
Using the expression for $h^2$ from equation(\ref{eqn:hsol1}), the evolution of the deceleration parameter is,
\begin{equation}
 q = -1 + \frac{1}{2} \left({3\eta (1+z)^3+{2(C_H -1)\over C_{\dot{H}}} (1-\eta) (1+z)^{2(C_H -1)/C_{\dot{H}}} \over \eta (1+z)^3 + (1-\eta)(1+z)^{2(C_H -1)/C_{\dot{H}}}} \right).
\end{equation}
In the limit $z>>1$ the second terms both on numerator and denominator inside the parenthesis on the right hand side of the above equation will be vanishingly 
small and be neglected, as a result, $q \to 1/2, $ confirming that the EDE behave like CDM in the earlier period\cite{}. More over at the special condition,
$3C_{\dot{H}}/2=C_H, \, \, \Omega_{m0}=1$ at which $\eta=1,$ the deceleration parameter become $q=1/2$ implies Einstein-de Sitter universe as we already noted 
in the previous section while dealing with the evolution of the Hubble parameter. As the universe evolves further the
 deceleration parameter 
deceases and become negative at the recent past in the evolution of the universe. The negative value of the deceleration parameter implies acceleration in the 
expansion of the universe. For small redshift, when the EDE would increasingly dominate, 
the deceleration parameter be stabilized at $q \to -1+(2C_H -1)/C_{\dot{H}},$ which is in fact 
less than $-1$ as per the best fit values of the parameters $C_H$ and $C_{\dot{H}}.$ This behavior 
of the parameter indicates a transition from an early decelerated epoch to a later accelerated phase of expansion, 
and this transition is found to be occurred at a redshift of $z_T=0.57.$
 Figure \ref{fig:decp} shows the behavior of the deceleration parameter with 
redshift.
The present value of the decelartion parameter is, $q(z=0)=-0.8.$ The observational constraints on the parameters, $q$ and $z_T$ are $q(z=0)=-0.64\pm 0.03$ 
\cite{Tegmark1,Hinshaw1}and $z_T = 0.45 - 0.73$\cite{Alam1}. Even though the predicted value of transition redshift $z_T$ in this model is almost in agreement with 
the observational constraint, 
the present value of the deceleration parameter is not in exact agreement but slightly less than the observational value.



\section{Big rip singularity}
\label{sec:estmation}
\begin{figure}
 \centering
\includegraphics[scale=0.7]{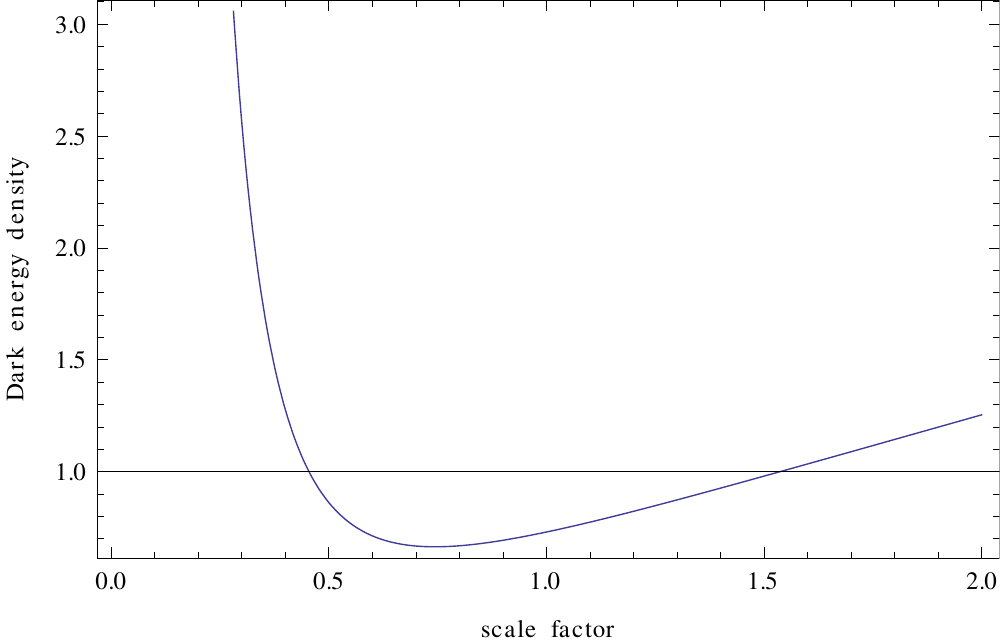}
\caption{Evolution of DE with scale factor. The plot shows that when the DE density is decreasing in the early stage but increasing at the later stage where the equation 
of state is less than -1.}
\label{fig:DEwitha}
\end{figure}

 In the earlier section we have seen that the EDE shows phantom nature for $z<0.257.$
During phantom phase dark energy density $\Omega_{EDE}$ will increases in proportion to the scale factor, see figure \ref{fig:DEwitha}. 
In this section we want to check whether the phantom nature of EDE indicates a 
big rip scenario. 
In the phantom stage the universe is increasingly DE dominated, hence the Hubble parameter
given in equation(\ref{eqn:hsol1}) can be approximated with the second term on the right side alone, 
\begin{equation}
 h^2 \simeq (1-\eta) \exp(-2(C_H -1)x/C_{\dot{H}} ).
\end{equation}
Considering the best fit values of the parameters, $C_H$ and $C_{\dot H}$, it can shown from the above equation that, 
the scale faactor will diverge in finite duration of time in future,
gives as big-rip time, $t_{rip},$
\begin{equation}
 t_{rip} \simeq  \frac{2}{(1-\eta)^{1/2}} H_0^{-1}.
\end{equation}
From the equation (\ref{eqn:eta}) of $\eta,$ the big rip time $t_{rip}$ is strongly depending on $\Omega_{m0}$ \cite{Caldwell1}. With 
$\Omega_{mo}=0.30,$  the above equation leads to $t_{rip}-t_0 \simeq 2.5 H_0^{-1}$ for the best estimates of the model parameters, and is around $36$ Giga Years for 
present model.
While EDE density behaves as  $\Omega{ede} \propto a$ in DE dominated phase, it also will diverge
\begin{figure}
 \centering
\includegraphics[scale=0.7]{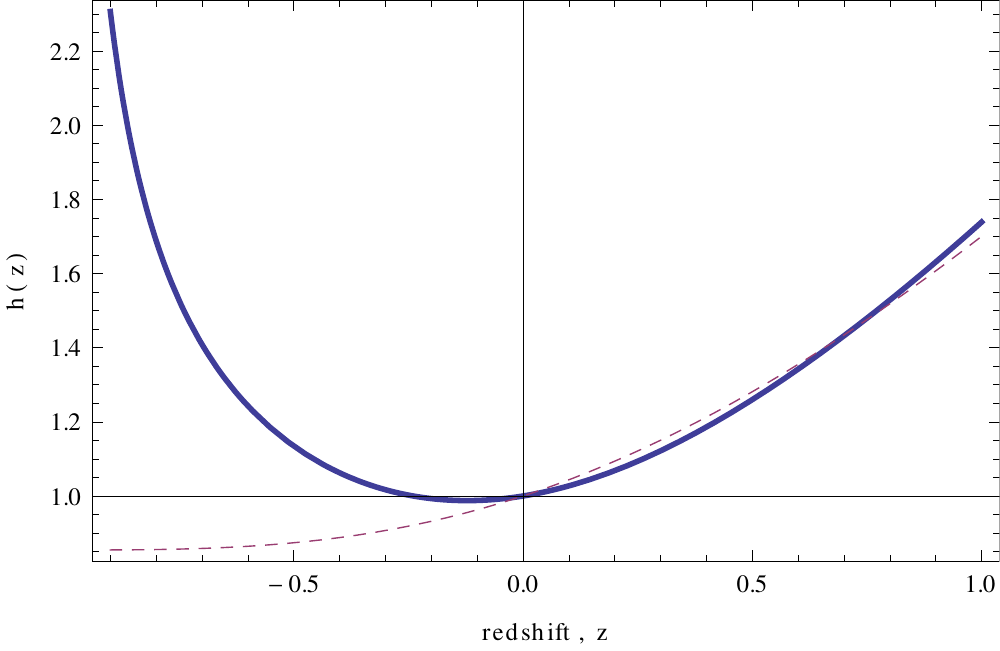}
\caption{Growth of the expansion rate especially for $z<0.$ Compared to $\Lambda$CDM model (the dotted line) the expansion is growing at a faster rate when $z<0$.}
\label{fig:h(z<0)}
\end{figure}
during the big rip. The big rip phenomenon due to the phantom nature of the dark energy was first discussed by Caldwell\cite{Caldwell1}.
The dark energy effectively anti gravitating in nature. In the phantom phase, the effect of this anti gravity is so high that it  
dissociates the large scale structures \cite{Caldwell1} in the universe. During this stage the  the expansion rate $H$ grows with time, 
see figure \ref{fig:h(z<0)}, so that the horizon (whose size is proportional to $H^{-1}$) 
closes in on as and the galaxies will thus crosses the horizon to disappear.

\section{Conclusions}
\label{sec:conclu} Entropic dark energy (EDE) was proposed first by Easson et al.\cite{Easson1} based on the Entropic gravity model. Technically 
the EDE arises from the surface term in Einstein-Hilbert action of gravity, and having a general form $C_H H^2+C_{\dot{H}} \dot{H}.$ Easson et al. have shown that the 
model predicts a transition form a decelerating phase to accelerating one at around a redshift, $z=0.5.$ They have also found a good agreement between the theoretical 
distance moduli of supernovae predicted by the model and the corresponding observational moduli.
We have analyzed the non-interacting EDE model, in which both the dark energy and dark matter satisfies seperate conservations laws, 
by evaluating the best estimates of the model parameters using the Union data on Type Ia supernovae.
After deriving the Hubble parameter, the evolutions of the dark energy density, equation of state parameter and deceleration parameter were studied. 
The model predicts an early decelerated epoch and a later accelerated epoch. The model will reduces to the Einstein-de Sitter universe for the 
special conditions of parameters,
$3C_{\dot{H}}/2=C_H$ and $\Omega_{m0} \simeq 1.$ 
The model predicts a slightly higher value for the mass parameter around, 
$\tilde{\Omega_{m0}} \simeq 1.25 \Omega_{m0},$ which may dilute the linear perturbation hence may negatively affect the structure formation. However a firm 
conclusion regarding this needs detailed calculation. The evolution of the deceleration parameter is studied and found that there is a transition form the 
early decelerated epoch to a later accelerated epoch at around a redshift $z=0.57,$ a value slightly higher than that predicted in the work of Easson at al., but well 
within the WMAP range\cite{Komatsu1}.

The equation of state is found crossing the phantom divide at $z<0.257$ and stabilizes at around 
$-1.3.$ It's present value is around $\omega_{EDE}\sim -1.185.$ 
We have also shown that the phantom nature of the EDE leads to big rip situation, such that during a finite time in the evolution of the universe the scale factor and 
dark energy density were become infinitely large. The big rip time is found to be around 36 Giga years from now. As pointed out by Caldwell, this may 
leads to the dissociation of structures and apart from that the horizon will approach us hence the structures may cross the horizon to disappear.

However one may the note the work of Spyros et al.\cite{Spyros1} at this juncture, where the authors treated the entropic dark energy as a dynamical vacuum with 
constant equation of state equal to -1, so that the dark energy 
is interacting with the dark matter sector and follows a common conservation law. In this work the authors have shown that the interacting EDE infact produce 
either eternal acceleration or deceleration, unless the dark energy density is modidifed with an additonal constant.

\section{On considering interaction between the cosmic components}

\noindent\textbf{ Acknowledgement}

One of the authors (TKM) is thankfull to IUCAA, Pune for the hospitality during the visit, where part of the work was done.

\end{document}